\documentclass[sigconf,screen]{acmart}

\acmSubmissionID{298}

\usepackage[inline]{enumitem}
\usepackage{pifont}
\usepackage{xcolor,colortbl}
\usepackage{acronym}
\usepackage{bm}
\usepackage[skip=1mm]{caption}
\usepackage{natbib}
\usepackage{hyperref}
\usepackage{amsmath}
\usepackage{amsfonts}

\usepackage{amssymb}
\hypersetup{
  colorlinks   = true,
  urlcolor     = blue,
  linkcolor    = blue,
  citecolor    = red
}
\usepackage{booktabs}
\usepackage{multirow}
\usepackage{siunitx}
\makeatletter
\let\algorithmic\@undefined
\let\endalgorithmic\@undefined
\makeatother

\usepackage{algorithm}
\usepackage{algpseudocode}
\usepackage{float}
\usepackage{cleveref}
\usepackage{amsmath}
\usepackage{subcaption}
\usepackage[normalem]{ulem}
\usepackage{tikz}
\usetikzlibrary{shapes, arrows.meta, positioning, matrix, calc, backgrounds, fit}
\usepackage{xcolor}
\usepackage{tikz}
\usepackage{tikz}
\usetikzlibrary{positioning} % REQUIRED for "below left=... of"
\usepackage{xcolor}          % usually already there, but safe

\usepackage{arydshln}

\definecolor{Gray}{gray}{0.85}
\definecolor{LightCyan}{rgb}{0.88,1,1}

\makeatletter
\newcommand\headingnodot{\def\@toclevel{4}%
  \@startsection{paragraph}{4}{\z@}%
  {-.2\baselineskip \@plus -2\p@ \@minus -.2\p@}%
  {-3.5\p@}%
  {\ACM@NRadjust{\bfseries}}}
\newcommand{\heading}[1]{\headingnodot{#1.}}
\makeatother

\AtBeginDocument{%
  }

\acrodef{DDRO}{direct document relevance optimization}
\acrodef{DSI}{differentiable search indexes}
\acrodef{GenIR}{generative information retrieval}
\acrodef{LM}{language model}
\acrodef{RLRF}{reinforcement learning from relevance feedback}
\acrodef{SFT}{supervised fine-tuning}

\newcommand{\NA}{\textemdash} % not available under inference-only reproducibility
\copyrightyear{2026}
\acmYear{2026}
\setcopyright{cc}
\setcctype{by}
\acmConference[SIGIR '26]{Proceedings of the 49th International ACM SIGIR Conference on Research and Development in Information Retrieval}{July 20--24, 2026}{Melbourne, VIC, Australia}
\acmBooktitle{Proceedings of the 49th International ACM SIGIR Conference on Research and Development in Information Retrieval (SIGIR '26), July 20--24, 2026, Melbourne, VIC, Australia}
\acmDOI{10.1145/3805712.3808567}
\acmISBN{979-8-4007-2599-9/2026/07}

\author{Kidist Amde Mekonnen}
\orcid{0009-0006-1702-9514}
\affiliation{
  \institution{University of Amsterdam}
\city{Amsterdam}
  \country{The Netherlands}
}
\email{k.a.mekonnen@uva.nl}

\author{Yongkang Li}
\orcid{0000-0001-6837-6184}
\affiliation{
  \institution{University of Amsterdam}
\city{Amsterdam}
  \country{The Netherlands}
}
\email{y.li7@uva.nl}

\author{Yubao Tang}
\orcid{0009-0003-8010-3404}
\affiliation{
  \institution{University of Amsterdam}
\city{Amsterdam}
  \country{The Netherlands}
}
\email{y.tang3@uva.nl}

\author{Simon Lupart }
\orcid{0009-0008-2383-4557}
\affiliation{
  \institution{University of Amsterdam}
\city{Amsterdam}
  \country{The Netherlands}
}
\email{s.c.lupart@uva.nl}

\author{Maarten de Rijke}
\orcid{0000-0002-1086-0202}
\affiliation{
 \institution{University of Amsterdam}
 \city{Amsterdam}
 \country{The Netherlands}
}
\email{m.derijke@uva.nl}
\begin{document}

% \title{Reproducing "Planning Ahead in Generative Retrieval" (PAG): Multilingual and Cross-Lingual Generalization, Robustness, and Scaling to Modern LLMs}
% \title{Planning in New Territories: A Reproducibility Study of Robustness, and Multilingual Generalization in Generative Retrieval}
% \title{Lost in Translation? Reproducing and Stress-Testing the Lexical Planner in Generative Retrieval}
% \title{Lost in Planning? Reproducing and Stress-Testing the Lexical Planner in Generative Retrieval}
% \title{Lost in Decoding? Reproducing and Stress-Testing the Lexical Planner in Generative Retrieval}
% \title{Lost in \emph{Planning}? Reproducing and Stress-Testing PAG’s Planning Signal in Generative Retrieval}
% \title{Lost in \emph{Planning}? Reproducing and Stress-Testing the Look-Ahead Signal in Generative Retrieval}
% \title{Lost in \emph{Planning}? Reproducing and Stress-Testing the Look-Ahead Prior in Generative Retrieval}

\title{Lost in \emph{Decoding}? Reproducing and Stress-Testing the Look-Ahead Prior in Generative Retrieval}

\begin{abstract}
Generative retrieval (GR) ranks documents by autoregressively generating document identifiers. Because many GR methods rely on trie-constrained beam search, they are vulnerable to early pruning of relevant prefixes under finite-beam decoding. \emph{Planning Ahead in Generative Retrieval} (PAG) mitigates this failure mode by using simultaneous decoding to compute a document-level look-ahead prior that guides subsequent sequential decoding.
We reproduce PAG at inference time and stress-test its decoding behavior. Using the authors' released checkpoint and identifier/trie artifacts under the reported decoding setup, we reproduce the main effectiveness results on MS~MARCO Dev and TREC-DL 2019/2020, and corroborate the reported beam-size--latency trade-off in our hardware setting. Beyond reproduction, we introduce \emph{plan drift} diagnostics that quantify how intent-preserving query variations, including misspellings, reordering, synonym substitutions, paraphrases, and naturality shifts, alter the planner's top-$n$ candidate set and highest-weight planner tokens, and how these changes affect guided decoding.
We find that PAG's planning signal is brittle under lexical surface-form variation: intent-preserving typos can trigger \emph{plan collapse}, where the planned candidate pool shifts enough that the look-ahead bonus provides little useful guidance, effectively reverting decoding toward weaker unguided search. We further evaluate fixed-index cross-lingual robustness using non-English \textsc{mMARCO} queries against an English index, and assess query-side mitigation strategies that require no re-indexing; query translation provides the strongest recovery in our setting.
Overall, our results confirm PAG's reported effectiveness and the benefit of planning-guided decoding under the released inference setup, while showing that these gains depend on the stability of the planning signal under realistic query variation and query--document mismatch. Code available at https://github.com/kidist-amde/lost-in-decoding.
\end{abstract}



\maketitle

\acresetall

\section{Introduction}
\label{sec:intro}
\textbf{Generative retrieval and prefix pruning.}
Generative retrieval (GR) reframes search as sequence generation: given a query, a model retrieves by autoregressively generating a document identifier (docid)~\cite{Tay2022DSI}.
At inference time, decoding is constrained to valid docids (e.g., via a trie), making retrieval sensitive to search-time errors.
In particular, beam search is myopic: it can discard globally relevant documents when their docid prefixes receive low probability early in generation.
This failure mode, \emph{prefix pruning}, occurs when a relevant document's prefix falls outside a beam of width $k$ and is no longer explored under finite-beam decoding~\cite{Zeng2024PlanningAI, stahlberg-byrne-2019-nmt, Constrained_mdr, jiang2026spend}.
Because decoding choices can dominate retrieval outcomes in GR, it is essential to verify whether reported gains persist under released artifacts and reported inference settings.
Moreover, for methods that rely on intermediate guidance signals, a further question arises: \emph{is the guidance itself stable under realistic query variation, or can it become a bottleneck?}

\heading{Intermediate signals for reliable decoding}
A growing line of GR work targets decoding reliability, motivated by the observation that end-to-end retrieval can be limited by \emph{search errors} rather than raw model capacity.
Prior work bridges generation and ranking by designing rank-aware identifiers, learning from relevance feedback (including reinforcement-style objectives), or directly optimizing document-level utility to better align token-level probabilities with retrieval quality~\citep{kumar2024surveygenir, Li2023LearningTR, mekonnen2025lightweight, zhou-etal-2023-enhancing-generative, ROGER, Constrained_mdr}.
Several approaches in GR and constrained decoding compute a discrete, cheaper intermediate signal (or a look-ahead estimate) to guide constrained search, and then refine predictions in a subsequent stage~\citep{Zeng2024PlanningAI, kumar2024surveygenir, li2025retrollm, qi2020prophetnet, tu-etal-2024-unlocking, nakshatri2025speculookahead, jiao2025look, lu-etal-2022-neurologic}.
The robustness of such intermediate signals under realistic input variation remains under-examined, even though instability can become a single point of failure in the retrieval pipeline~\citep{liu2024robustneuralIR, kumar2024surveygenir, li_attck,Li26_ecir}.
Since real retrieval environments are noisy due to misspellings, paraphrases, segmentation ambiguity, and productive morphology, we argue for evaluations that go beyond end metrics to \emph{instrument} the intermediate signal itself, separating cases where it provides consistent look-ahead guidance from cases where it destabilizes decoding and amplifies search errors.

\heading{Planning Ahead in Generative Retrieval (PAG)}
Planning Ahead in Generative Retrieval (PAG)~\citep{Zeng2024PlanningAI} proposes a two-stage decoding strategy to mitigate \emph{prefix pruning} in trie-constrained beam search.
It first derives a fast, query-dependent planning signal via \emph{simultaneous decoding} and uses it to score and shortlist candidate documents.
It then uses these planning scores as a \emph{look-ahead bonus} during constrained decoding, favoring prefixes supported by high-scoring planned documents and reducing harmful early pruning.

\heading{Why reproduce and stress-test PAG?}
Reproduction is particularly valuable for PAG because it targets a central GR failure mode: prefix pruning under trie-constrained beam search.
\begin{enumerate*}[label=(\roman*)]
\item PAG reports improved effectiveness--efficiency trade-offs, achieving strong retrieval with smaller beams by using a single-pass planning signal as a look-ahead bonus during constrained decoding; verifying these gains using the released artifacts and the reported decoding configuration is important for assessing practical utility.
\item Because the look-ahead bonus is computed from a top-$n$ planning set, guided decoding depends on that set's coverage: prefixes unsupported by planned documents receive no bonus, making the method directly vulnerable to query variation and distribution shift.
\item GR pipelines involve expensive corpus-side construction (e.g., identifier assignment and trie indexing), motivating tests of whether query-side shift that harms planning coverage or alignment can be mitigated without rebuilding the index.
\end{enumerate*}
Finally, aggregate metrics can mask \emph{tail risk}: when the planning set omits relevant documents, the look-ahead bonus cannot favor their docid prefixes, making recovery under finite-beam decoding less likely.
% We therefore instrument the planner with candidate-coverage and drift diagnostics to characterize when planning provides reliable guidance and how reliability degrades under shift.
We therefore instrument the planner with overlap and drift-based diagnostics to characterize when planning provides reliable guidance and how reliability degrades under shift.

\heading{Stress tests and scope}
We conduct an inference-time reproduction and two stress tests of PAG.
% First, we evaluate robustness under intent-preserving query variations and translation-based variants.
First, we evaluate robustness under intent-preserving query variations.
Second, we evaluate a stricter \emph{query--document language mismatch} setting by issuing non-English \textsc{mMARCO} queries~\cite{Bonifacio2021mMARCOAM} against the fixed English \textsc{MS MARCO} collection and released identifier trie, without re-indexing.
This mismatch setting is a direct test of whether PAG's planning mechanism remains useful when query-side surface form diverges from the evidence space on which the planner and identifier trie were built.
For this setting, we evaluate two query-side mitigations with corpus-side artifacts fixed: (i) query-only translation and (ii) trained query-side adaptation via planner-token distillation from aligned English queries.

Our study is organized around three research questions:
\begin{enumerate}[leftmargin=*,label=\textbf{RQ\arabic*},nosep]
\item \textbf{Inference-time validation:} To what extent can we reproduce PAG's reported effectiveness and inference-time analyses measurable \emph{without retraining} under the released artifacts and decoding configuration, and what effectiveness--efficiency trends emerge across beam sizes?
% \item \textbf{Robustness \& plan drift:} How does intent-preserving query variation affect planning stability, candidate coverage, and downstream ranking, relative to strong dense and GR baselines?
\item \textbf{Robustness \& plan drift:} How does intent-preserving query variation affect planning stability, planned-set overlap, and downstream ranking, relative to strong dense and GR baselines?
\item \textbf{Cross-lingual shift with a fixed English index:} Under language mismatch with a fixed English identifier trie, how much performance can be recovered without re-indexing via (a) zero-shot use, (b) query-only translation, and (c) planner-token distillation?
\end{enumerate}

\heading{Contributions}
\begin{enumerate*}[label=(\roman*)]
\item We reproduce PAG's inference-time results on MS~MARCO Dev and TREC-DL 2019/2020 under the released checkpoint, identifiers, trie, and reported decoding configuration, validating the reported effectiveness and characterizing inference-time behavior under our hardware setup.
\item We introduce \emph{plan drift} as instability in the planner's top-$n$ candidate set and high-weight planner tokens under intent-preserving query variation, and show how this instability weakens, misaligns, or in some cases collapses planning-guided decoding.
\item We stress-test PAG under query variation and fixed-index cross-lingual query shift, showing that restoring query-side compatibility through translation is substantially more effective than lightweight planner-token alignment without re-indexing.\footnote{Following ACM's terminology, our study is a combination of an artifact-based \emph{reproducibility study} (for RQ1, we follow the ``different team, same setup'' mode, as we re-execute released artifacts under the reported setup, without retraining) and a \emph{replicability study} (for RQ2 and RQ3, we follow the ``different team, different setup'' mode as we conduct stress-tests with query variations and cross-lingual query shifts).}
\end{enumerate*}

\section{Related Work}
\label{sec:related}

\heading{Generative retrieval, constrained decoding, and guidance}
Generative retrieval (GR) ranks documents by generating corpus-specific identifiers (docids) rather than scoring documents directly in an embedding space~\citep{Tay2022DSI,Bevilacqua2022SEAL,NCI,tang2024recentWWW,tang2023recent}.
Because decoding is constrained to valid identifiers (e.g., via a trie), retrieval can be limited by search errors, including \emph{prefix pruning} under finite-beam decoding~\citep{stahlberg-byrne-2019-nmt,pradeep2023scale, Constrained_mdr, jiang2026spend}.
A broad response is to augment constrained decoding with auxiliary guidance or look-ahead estimates~\citep{Zeng2024PlanningAI, qi2020prophetnet, li2025retrollm, tu-etal-2024-unlocking, nakshatri2025speculookahead, jiao2025look, lu-etal-2022-neurologic}.
Within GR, guidance often combines multiple signals (e.g., lexical/semantic hybrids, ranking-oriented objectives, or alternative generation procedures) to improve decoding reliability~\citep{kuzi2020leveraging, kumar2024surveygenir, zhou2022ultron, zhou-etal-2023-enhancing-generative, mekonnen2025lightweight, Li2023LearningTR,10.1145/3626772.3661379,tang2024listwise, dong2026multi, ROGER}.
PAG instantiates this line of work by using a fast planning-derived look-ahead bonus to bias trie-constrained decoding~\citep{Zeng2024PlanningAI}.
In contrast to proposing a new guidance mechanism, we study whether PAG's released planning signal remains reliable under realistic query variation and fixed-index query--document mismatch.
\heading{Robustness to query variation and intermediate-signal stability}
Intent-preserving query variation (e.g., typos, paraphrases, and reordering) is a standard retrieval stress test and has been operationalized through query-variation generators and UQV-style taxonomies~\citep{penha_qvg,hagen-etal-2024-revisiting,liu2025robustness}.
Such variation can be challenging for pipelines that rely on discrete intermediate predictions or guidance signals, since small surface changes may shift early-stage outputs and propagate downstream~\citep{liu2024robustneuralIR,li_attck}.
Most robustness evaluations emphasize end-to-end effectiveness~\citep{liu2025robustness,FGSM,li2025reproducing}, with less attention to the stability of intermediate components that steer search.
Our \emph{plan drift} analysis addresses this gap for PAG by quantifying changes in the planner's candidate set and high-weight planner tokens, and relating these changes to downstream ranking behavior.

\heading{Cross-lingual retrieval and query--document mismatch}
Cross-lingual retrieval increases query--document mismatch and can weaken methods that rely on surface-form overlap; for example, \citet{Bonifacio2021mMARCOAM} show that translation artifacts can substantially degrade lexical baselines relative to dense models.
While cross-lingual dense retrieval is well studied~\citep{nair2022transfer}, GR is still commonly evaluated in monolingual settings~\citep{kumar2024surveygenir}, and multilingual GR efforts have mainly focused on multilingual identifier learning and compression~\citep{zhang2024multilingual}.
We instead study PAG under fixed-index query--document language mismatch, where the English identifier space and document-side artifacts remain unchanged, and evaluate query-side mitigations that do not require re-indexing.
\section{Reproducibility Methodology}
\label{sec:methodology}

This section formalizes the reproduced PAG inference pipeline and introduces the diagnostics used in our robustness analyses. Table~\ref{tab:notation} summarizes the main notation used throughout the section.

\subsection{Problem Formulation}
\label{subsec:formulation}

A GR model ranks documents by autoregressively decoding length-$L$ docids
$c_d=[c_{d,1},\ldots,c_{d,L}]$ conditioned on a query $q$.
Inference uses trie-constrained beam search over valid prefixes.
Decoding is scored additively using decoder states $h_i(c_{<i},q)$ and step-specific docid-token embeddings $E_i[\cdot]$.
A complete identifier is scored by
\begin{equation}
s(c_d;q)=\sum_{i=1}^{L} E_i[c_{d,i}] \cdot h_i(c_{d,<i},q)
\label{eq:seq_score}
\end{equation}
For a prefix $c_{\le i}$, the corresponding prefix score can be written recursively as
\begin{equation}
s(c_{\le i};q)= s(c_{<i};q) + E_i[c_i]\cdot h_i(c_{<i},q),
\qquad s(c_{\le 0};q)=0,
\label{eq:prefix_score}
\end{equation}
which is equivalent to $s(c_{\le i};q)=\sum_{j=1}^{i} E_j[c_j]\cdot h_j(c_{<j},q)$.
Trie-constrained beam search retains the top-$k$ \emph{valid} prefixes at each step and expands them only along trie edges.\footnote{Equivalently, validity can be enforced by adding a mask $g(c_{\le i})$, with $g=0$ for valid prefixes and $g=-\infty$ otherwise.}
A key failure mode is \emph{prefix pruning}: once a relevant docid prefix drops out of the beam, trie constraints prevent it from being revisited, even if completing it would yield a highly relevant document.
This motivates decoding strategies that incorporate document-level look-ahead guidance.

\begin{table}[t]
\small
\caption{Key notation used throughout the reproduced PAG pipeline and our robustness diagnostics.}
\label{tab:notation}
\centering
\begin{tabular}{@{}ll@{}}
\toprule
Symbol & Meaning \\
\midrule
$c_d$ & Sequential docid of document $d$ (length $L$) \\
$t_d$ & Set-based planning identifier of document $d$ ($m$ tokens) \\
$s(c_{\le i}; q)$ & Sequential prefix score (Eq.~\ref{eq:prefix_score}) \\
$s_{\mathrm{simul}}(q,d)$ & Simultaneous planning score (Eq.~\ref{eq:simul_score}) \\
$b(c_{\le i})$ & Look-ahead bonus from planned documents \\
$D$ / $D_n(q)$ & Top-$n$ planning set (default $n{=}1000$) \\
$k$ & Beam size (default $100$) \\
$m$ & Planning tokens per document (default $64$) \\
$K$, $\ell$ & Truncation depths for overlap diagnostics (default $100$) \\
$\tau$, $\delta$ & Stability and drop thresholds for collapse \\
\bottomrule
\end{tabular}
\end{table}

%__________________________________________________
\subsection{Planning Ahead in Generative Retrieval}
\label{subsec:pag}

\begin{figure}[t]
  \centering
  \includegraphics[width=\columnwidth]{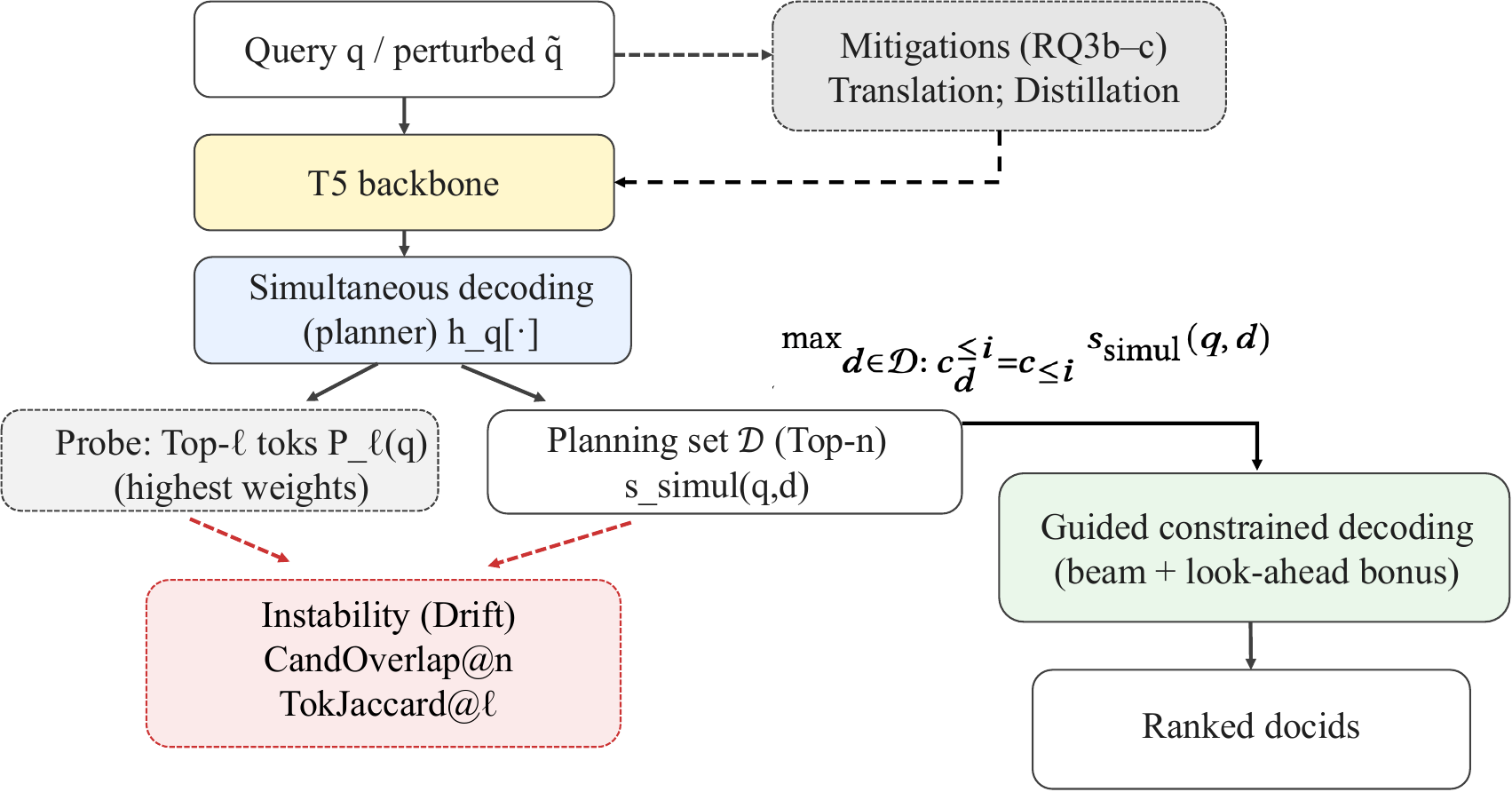}
% \caption{\textbf{Compact PAG pipeline with probes.} 
% Simultaneous decoding produces planner weights $h_q[\cdot]$, inducing a top-$n$ planning set $\mathcal{D}$ used to compute a look-ahead bonus during trie-constrained decoding. 
% \textbf{Red components} denote our reproducibility diagnostics: planner instability is probed via candidate drift (CandOverlap@{$100$}) and token drift (TokJaccard@{$\ell$}). 
% Dashed box (top right) denotes language-shift mitigations (RQ3): query translation and query-side planner alignment via distillation, keeping corpus-side artifacts fixed.}
\caption{\textbf{Compact PAG pipeline with probes.}
Simultaneous decoding yields planner weights $h_q[\cdot]$, inducing a top-$n$ candidate pool used to compute a look-ahead bonus during trie-constrained decoding.
\textbf{Red} marks our diagnostics: candidate drift (CandOverlap@100) and token drift (TokJaccard@100).
The top-right dashed box summarizes RQ3 query-side mitigations (translation and planner alignment) with corpus-side artifacts fixed.}

  \label{fig:pag_pipeline_probes}
  \vspace{-1mm}
\end{figure}

PAG~\cite{Zeng2024PlanningAI} reduces prefix pruning by pairing each document's sequential docid $c_d$ with a set-based docid
$t_d=\{t_{d,1},\ldots,t_{d,m}\}$, an unordered set of $m$ planning tokens drawn from a planning vocabulary.
At inference time, PAG combines a fast planning step with planning-guided constrained decoding. Figure~\ref{fig:pag_pipeline_probes} summarizes the reproduced PAG pipeline and our probes.

\heading{Simultaneous decoding (planning)}
Given $q$, one-step simultaneous decoding produces query-dependent token weights $h_q[\cdot]$ over the planning vocabulary.
Documents are scored by aggregating weights over $t_d$:
\begin{equation}
s_{\text{simul}}(q,d)=\sum_{j=1}^{m} h_q[t_{d,j}],
\label{eq:simul_score}
\end{equation}
and the top-$n$ documents under $s_{\text{simul}}$ form the planning set $D$.

\heading{Planning-guided constrained decoding}
In trie-constrained beam search over $c_d$, each valid prefix $c_{\le i}$ receives a look-ahead bonus from compatible planned documents.
Let
\begin{equation}
D_{c_{\le i}}=\{d\in D : c_{d,\le i}=c_{\le i}\},
\label{eq:compatible_set}
\end{equation}
and define
\begin{equation}
b(c_{\le i})=
\begin{cases}
\max\limits_{d \in D_{c_{\le i}}} s_{\text{simul}}(q,d) & \text{if } D_{c_{\le i}}\neq \emptyset,\\
0 & \text{otherwise.}
\end{cases}
\label{eq:bonus}
\end{equation}
PAG scores prefixes for pruning as
\begin{equation}
s'(c_{\le i}; q)=
\underbrace{s(c_{\le i}; q)}_{\text{sequential prefix score}}
+
\underbrace{b(c_{\le i})}_{\text{planning look-ahead bonus}}
\label{eq:pag_score}
\end{equation}
Beam pruning ranks prefixes using $s'(c_{\le i};q)$, while expansions remain trie-constrained and add the next-token sequential contribution in Eq.~\eqref{eq:prefix_score}.
This promotes prefixes that can still complete to highly scored planned documents, reducing early pruning under finite beams.

%__________________________________________________

\subsection{PAG Optimization Pipeline}
\label{subsec:pag_training}

PAG trains a single backbone to support (i) set-based planning scores and (ii) sequential docid decoding, via three stages
(see~\cite{Zeng2024PlanningAI} for full objectives and sampling).

\begin{itemize}[leftmargin=*]
\item \textbf{Stage 1 (set-based planning).}
Learn a sparse lexical planner in two steps:
(i) train a sparse encoder $M_{\text{sp}}$ with a MarginMSE objective and a FLOPs regularizer to produce document token
weights $w_d\in\mathbb{R}^{V_T}$;
(ii) define set-docids by top-$m$ selection $t_d \gets \mathrm{Top}\text{-}m(w_d)$, then fine-tune a set model $M_{\text{set}}$
with $L_{\text{set}}$ so that query weights $h_q[\cdot]$ yield high
$s_{\text{simul}}(q,d)=\sum_{t\in t_d} h_q[t]$ for relevant documents.

\item \textbf{Stage 2 (sequential decoding).}
Build semantic sequential docids via residual quantization:
$c_d \gets \mathrm{RQ}(d)\in\mathcal{V}^L$.
Train a sequential model $M_{\text{seq}}$ as a generative retriever over $c_d$ under trie constraints, using the
prefix-oriented loss $L_{\text{seq}}$ (supervising intermediate prefixes $c_{d,\le i}$ for $i=1,\dots,L$) to reduce
prefix pruning during constrained decoding.

\item \textbf{Stage 3 (unified model).}
Combine both capabilities in one model by initializing
$M \leftarrow \mathrm{Avg}(M_{\text{set}},M_{\text{seq}})$ (parameter averaging), then fine-tuning with the joint objective
$L_{\text{set}}+L_{\text{seq}}$.
To keep simultaneous decoding compatible with the unified backbone, the decoder is additionally conditioned on query
tokens while preserving trie-constrained decoding for sequential docids.
\end{itemize}
%__________________________________________________

\subsection{Plan Stability, Plan Sensitivity, and Plan Collapse}
\label{subsec:metrics}

For each query $q$, PAG’s planning stage produces a top-$n$ candidate set $D_n(q)$ (top-$n$ documents under
$s_{\text{simul}}(q,d)$) and a planner token set $P_\ell(q)$, the top-$\ell$ planning-vocabulary tokens under query weights
$h_q[\cdot]$. Given an original query $q$ and an intent-preserving variation $\tilde q$, we quantify plan stability and
sensitivity via overlap between $D_K(q)$ and $D_K(\tilde q)$ and between $P_\ell(q)$ and $P_\ell(\tilde q)$, and report
plan collapse rates under a thresholded criterion. Unless otherwise stated, we use $K=100$ and $\ell=100$ for overlap
diagnostics, and set $n=1000$ as the default Stage-1 candidate-pool size used during inference, following PAG’s reported
setting. Table~\ref{tab:diagnostics} summarizes the plan-drift diagnostics defined in this subsection.

\heading{Candidate-set stability}
Let $D_K(q)$ denote the top-$K$ truncation of the Stage-1 candidate pool $D_n(q)$ (with $K \ll n$), and let
$I_K(q,\tilde q)=|D_K(q)\cap D_K(\tilde q)|$. We report:
\begin{equation}
\mathrm{CandOverlap}@K(q,\tilde q)=\tfrac{I_K(q,\tilde q)}{K}
\label{eq:candoverlap}
\end{equation}

\heading{Planner-token stability}
Let $J_\ell(q,\tilde q)=|P_\ell(q)\cap P_\ell(\tilde q)|$. We report token-set Jaccard similarity:
\begin{equation}
\mathrm{TokJaccard}@\ell(q,\tilde q)=\tfrac{J_\ell(q,\tilde q)}{|P_\ell(q)\cup P_\ell(\tilde q)|}
\label{eq:tokjaccard}
\end{equation}
We compute these per query and summarize them using mean, median, and tail quantiles (p10, p25, p75, p90).

\heading{Planning-only retrieval (\textsc{SimulOnly})}
Let $M(\cdot)$ denote the evaluation metric (MRR@10 for MS~MARCO Dev, NDCG@10 for TREC-DL). \textsc{SimulOnly} ranks
documents by $s_{\text{simul}}(q,d)$; we report the metric change under variation:
\begin{equation}
\Delta M_{\text{SimulOnly}}(q,\tilde q)=M_{\text{SimulOnly}}(\tilde q)-M_{\text{SimulOnly}}(q)
\label{eq:deltasimul}
\end{equation}

\heading{Plan sensitivity (counterfactual plan swap)}
To isolate sensitivity to look-ahead guidance from the intrinsic difficulty of $\tilde q$, we decode $\tilde q$ twice under
identical sequential decoding and trie constraints: (i) with its own plan (\emph{normal}) and (ii) using the clean-query plan
computed for $q$ (\emph{swapped}). This is a counterfactual diagnostic rather than a retrieval setting: it reuses a planning signal from a different query to isolate the effect of guidance quality.
\begin{equation}
% PLEASE LEAVE THIS HERE SO THAT WE AVOID THAT THE EQUATION NUMBER JUMPS TO THE NEXT LINE
\mbox{}\hspace*{-1mm}
\mathrm{PlanSwapDrop}(q,\tilde q)\!=\!M_{\text{PAG}}(\tilde q;\text{normal})-M_{\text{PAG}}(\tilde q;\text{swapped})
% AGAIN, PLEASE LEAVE THIS HERE SO THAT WE AVOID THAT THE EQUATION NUMBER JUMPS TO THE NEXT LINE
\hspace*{-1mm}\mbox{}
\label{eq:planswapdrop}
\end{equation}
Negative values indicate that the clean-query plan improves effectiveness on $\tilde q$ (i.e., the perturbed plan is harmful).

\heading{Guided decoding gain over planning}
We report the marginal gain of full PAG over planning-only retrieval on $\tilde q$:
\begin{equation}
\mathrm{SeqGain}(\tilde q)=M_{\text{PAG}}(\tilde q)-M_{\text{SimulOnly}}(\tilde q)
\label{eq:seqgain}
\end{equation}

\heading{Low-stability tail events (``plan collapse'')}
We define \emph{plan collapse} as a query-level tail event where planner stability is low and planning-only effectiveness drops sharply. Using a percentile-based stability threshold $\tau$ and an effectiveness-drop threshold $\delta$, a query is flagged as collapsed if
\begin{equation}
\label{eq:collapse}
\begin{aligned}
\bigl(\mathrm{CandOverlap}@K < \tau \;\lor\; \mathrm{TokJaccard}@\ell < \tau\bigr) \\
\land\; \Delta M_{\text{SimulOnly}}(q,\tilde q) \le -\delta
\end{aligned}
\end{equation}
Unless otherwise stated, $\tau$ is computed per condition (split $\times$ variation $\times$ seed) as the 10th percentile of the $\mathrm{CandOverlap}@K$ distribution. We report sensitivity by sweeping the $\tau$-percentile and the absolute-drop threshold $\delta$, and we additionally report lower-tail quantiles (e.g., p10/p25) of the stability metrics.

\begin{table}[t]
\footnotesize
\caption{Plan-drift diagnostics at a glance.}
\label{tab:diagnostics}
\centering
\begin{tabular}{@{}lp{0.58\columnwidth}@{}}
\toprule
Diagnostic & What it measures \\
\midrule
CandOverlap@$K$ & Stability of the top-$K$ planned candidate set \\
TokJaccard@$\ell$ & Stability of the top-$\ell$ planner-token set \\
$\Delta M_{\mathrm{SimulOnly}}$ & Change in planning-only effectiveness under variation \\
PlanSwapDrop & Effect of using the perturbed plan rather than the clean plan \\
SeqGain & Gain of full PAG over planning-only retrieval \\
\bottomrule
\end{tabular}
\end{table}

\section{Experimental Setup}
\label{sec:setup}

\textbf{Reproducibility scope.}
Our experiments rely on the authors’ released checkpoint and corpus-side artifacts (docids and trie). When intermediate artifacts needed to reconstruct a training stage are unavailable, we treat the corresponding components as fixed and explicitly mark results that would require retraining to reproduce. Unless stated otherwise, RQ1--RQ2 use the released PAG artifacts under the reported inference-time decoding configuration, while RQ3 evaluates fixed-index cross-lingual query shift and is reported separately from artifact-level reproduction claims.

\subsection{Released artifacts and experimental scopes}
\label{subsec:scope_artifacts}

\heading{Inference-time reproducibility (RQ1, RQ2)}
We evaluate PAG using the released T5-base checkpoint, the trie built over 8.8M sequential passage identifiers ($L=8$, $V=2048$), and the stored set-based identifiers used for simultaneous planning scores $s_{\text{simul}}(q,d)$. We follow the reported decoding procedure without modifying document identifiers.

\heading{Cross-lingual query shift (RQ3)}
We issue non-English \textsc{mMARCO} queries against the fixed English MS~MARCO passage collection using the same released docids and trie. This setting evaluates query--corpus language mismatch without re-indexing. As query-side mitigations, we evaluate translation into English before planning and decoding, and a learned planner-alignment model. For translation, we use M2M100~\cite{M2M100}.\footnote{Implementation via the
Transformers M2M100 model documentation: \url{https://huggingface.co/docs/transformers/en/model_doc/m2m_100}.}

\subsection{Datasets and metrics}
\label{subsec:data}
We use the MS~MARCO passage retrieval benchmark (8.8M passages) and report results on MS~MARCO Dev (6,980 queries) and TREC-DL 2019/2020. Following the original paper, we report MRR@10 on MS~MARCO Dev, NDCG@10 on TREC-DL, and Recall@10 on all datasets. For RQ3, we use non-English \textsc{mMARCO} query sets while keeping the document corpus and relevance judgments fixed to the English MS~MARCO passage collection.

\begin{table*}[t]
\centering
% \caption{PAG effectiveness reproduction (cf.\ Table~1 in the original paper) with additional recent dense baselines for context.
% $^{\uparrow}$/$^{\downarrow}$: significantly higher/lower than PAG (t-test with Bonferroni correction, $p<0.01$). Dense retrieval uses brute-force scoring over the full corpus. We explicitly separate baselines reported in the original PAG work (top) from additional modern dense retrievers we ran for broader context (bottom). All additional baselines use brute-force scoring.}
\caption{PAG effectiveness reproduction (cf.\ original Table~1) with dense-retrieval baselines for context.
$^{\uparrow}$/$^{\downarrow}$ indicate significantly higher/lower than reproduced PAG (t-test with Bonferroni correction, $p<0.01$).
Dense baselines are scored by brute-force over the full corpus; we separate baselines reported in~\cite{Zeng2024PlanningAI} from our additional runs.}
\label{tab:rq1_effectiveness}
\setlength{\tabcolsep}{4pt}
\begin{tabular}{lcc cc cc cc}
\toprule
\textbf{Method} & \textbf{KD} & \textbf{Index Mem. (GB)} &
\multicolumn{2}{c}{\textbf{MS~MARCO Dev}} &
\multicolumn{2}{c}{\textbf{TREC-DL 2019}} &
\multicolumn{2}{c}{\textbf{TREC-DL 2020}} \\
\cmidrule(r){4-5}
\cmidrule(r){6-7}
\cmidrule(r){8-9}
& & & MRR@10 & Recall@10 & NDCG@10 & Recall@10 & NDCG@10 & Recall@10 \\
\midrule
\multicolumn{9}{l}{\emph{Dense retrieval baselines reported in~\cite{Zeng2024PlanningAI} (for reference)}} \\
TCT-ColBERT & \ding{51} & 25.30 & 0.335\rlap{$^{\downarrow}$} & 0.596\rlap{$^{\downarrow}$} & 0.670\rlap{$^{\downarrow}$} & 0.240\rlap{$^{\downarrow}$} & 0.668\rlap{$^{\downarrow}$} & 0.218\rlap{$^{\downarrow}$} \\
MarginMSE   & \ding{51} & 25.30 & 0.325\rlap{$^{\downarrow}$} & 0.581\rlap{$^{\downarrow}$} & 0.699\rlap{$^{\downarrow}$} & 0.250\rlap{$^{\downarrow}$} & 0.645\rlap{$^{\downarrow}$} & 0.203\rlap{$^{\downarrow}$} \\
TAS-B       & \ding{51} & 25.30 & 0.344\rlap{$^{\downarrow}$} & 0.622\rlap{$^{\downarrow}$} & 0.717\rlap{$^{\uparrow}$}  & 0.255\rlap{$^{\downarrow}$} & 0.685\rlap{$^{\downarrow}$} & 0.230\rlap{$^{\downarrow}$} \\
CL-DRD      & \ding{51} & 25.30 & 0.382\rlap{$^{\downarrow}$} & 0.651\rlap{$^{\downarrow}$} & 0.725\rlap{$^{\uparrow}$}  & 0.266               & 0.687\rlap{$^{\downarrow}$} & 0.216\rlap{$^{\downarrow}$} \\
\hdashline
\multicolumn{9}{l}{\emph{Additional dense retrievers (our runs; contextual reference only)}} \\
% Contriever  & \ding{55} & 25.30 & 0.341\rlap{$^{\downarrow}$} & 0.627\rlap{$^{\downarrow}$} & 0.676\rlap{$^{\downarrow}$} & 0.169\rlap{$^{\downarrow}$} & 0.667\rlap{$^{\downarrow}$} & 0.232\rlap{$^{\downarrow}$} \\
Nomic-v2    & \ding{55} &25.30 & 0.341\rlap{$^{\downarrow}$} & 0.616\rlap{$^{\downarrow}$} &
0.705  & 
0.154\rlap{$^{\downarrow}$}&
0.684\rlap{$^{\downarrow}$} & 
0.227\rlap{$^{\downarrow}$}  \\

EmbeddingGemma & \ding{55} & 25.30 & 0.325\rlap{$^{\downarrow}$} & 0.591\rlap{$^{\downarrow}$} & 0.730\rlap{$^{\uparrow}$}  & 0.168\rlap{$^{\downarrow}$} & 0.721\rlap{$^{\uparrow}$}  & 0.231\rlap{$^{\downarrow}$} \\
Qwen3-8B    & \ding{55} & 135.10 & 0.368\rlap{$^{\downarrow}$} & 0.670 & 0.763\rlap{$^{\uparrow}$}  & 0.178\rlap{$^{\downarrow}$} & 0.743\rlap{$^{\uparrow}$}  & 0.249\rlap{$^{\uparrow}$} \\
\midrule
\multicolumn{9}{l}{\emph{Generative retrieval baselines}} \\
NOVO        & \ding{55} & 0.80   & 0.126 & 0.242 & 0.258 & 0.112 & 0.310 & 0.140 \\
MINDER      & \ding{55} & 12.16  & 0.186 & 0.383 & 0.506 & 0.201 & 0.392 & 0.144 \\
LTRGR       & \ding{55} & 12.16  & 0.255 & 0.531 & 0.598 & 0.238 & 0.553 & 0.182 \\
RIPOR       & \ding{51} & 1.06   & 0.333 & 0.562 & 0.628 & 0.205 & 0.631 & 0.191 \\
\midrule
PAG (reported)   & \ding{51} & 3.27 & 0.385 & 0.670 & 0.705 & 0.267 & 0.700 & 0.236 \\
PAG (reproduced) & \ding{51} & 3.27 & 0.386 & 0.671 & 0.703 & 0.265 & 0.701 & 0.236 \\
\bottomrule
\end{tabular}
\end{table*}

\subsection{Model and identifier configuration}
\label{subsec:model}

Experiments use the released T5-base checkpoint and fixed corpus-side artifacts. Each document $d$ has (i) a sequential docid $c_d=[c_{d,1},\ldots,c_{d,L}]$ constructed via residual quantization (RQ) with $L=8$ and docid-token vocabulary size $V=2048$, and (ii) a set-based identifier $t_d=\{t_{d,1},\ldots,t_{d,m}\}$ with $m=64$ planning tokens.

\subsection{Inference and decoding settings}
\label{subsec:decode}
Retrieval uses trie-constrained beam search with the planning-guided prefix score in Eq.~\ref{eq:pag_score}. Unless otherwise stated, inference follows the paper’s default configuration: beam size $k=100$ and a planning set formed by the top $n=1000$ documents under $s_{\text{simul}}(q,d)$.

\subsection{Query variations}
\label{subsec:variation}

For RQ2, we generate perturbed queries offline and keep them fixed across runs. Variations follow the UQV taxonomy~\citep{penha_qvg, li2025reproducing}: \emph{misspelling}, \emph{reordering}, \emph{synonym}, \emph{paraphrase}, and \emph{naturality}. We instantiate variations with five seeds (1999, 5, 27, 2016, 2026) and report mean$\pm$std across the resulting variation sets.

\subsection{Cross-lingual query shift and adaptations}
\label{subsec:crosslingual}

For RQ3, we evaluate four languages (Chinese, Dutch, French, and German) using released \textsc{mMARCO} query sets.\footnote{\url{https://huggingface.co/datasets/unicamp-dl/mmarco}.}
These languages induce different sources of mismatch with the English planning vocabulary: Chinese introduces a script mismatch; German and Dutch exhibit richer morphology; French shares script but differs lexically.

\heading{RQ3 baselines (inference-only)}
We compare:
\begin{enumerate}[leftmargin=*]
  \item \textbf{Naive cross-lingual PAG:} apply the released English PAG pipeline directly to non-English queries.
  \item \textbf{Sequential-only:} disable the planning bonus and decode using only the autoregressive score $s(c_{\le i};q)$.
  \item \textbf{Translate-at-inference:} translate non-English queries into English, then run the unmodified English PAG pipeline.
  % \item \textbf{Dense multilingual baseline:} apply a modern multilingual dense retriever to the same fixed English corpus for external calibration under cross-lingual query shift.
\end{enumerate}

\heading{Query-side adaptation: planner alignment (no re-indexing)}
As a lightweight learned mitigation, we align non-English queries to the released English planner using paired \textsc{mMARCO} queries with the same query id. Let $q^{\text{en}}$ be an English query and $q^{\ell}$ its paired non-English query for $\ell\in\{\texttt{nl},\texttt{fr},\texttt{de},\texttt{zh}\}$. We use the released English PAG checkpoint as a frozen teacher and initialize a student from the same checkpoint; training qids are disjoint from evaluation, which uses the same dev qids as our inference-only baselines. All document-side artifacts remain fixed ($t_d$, $c_d$, trie, and released index files), i.e., no re-indexing.

For each pair, the teacher is scored on $q^{\text{en}}$ and the student on $q^{\ell}$ over the \emph{English planning vocabulary}. Concretely, planner scores for vocabulary token $v$ are
\[
z[v]=\max_{t}\Big(\log(1+\mathrm{ReLU}(\texttt{lexical\_logit}_{t}[v]))\cdot m_t\Big),
\]
where $m_t$ is the attention mask. We update only query-side student parameters (encoder and \texttt{lm\_head}), while keeping the decoder frozen.

We optimize temperature-scaled distillation:
\begin{equation}
\mathcal{L}_{\text{align}}
=
\tau^2\,
\mathrm{KL}\!\left(
\mathrm{softmax}\!\left(\tfrac{z_{\text{teach}}}{\tau}\right)\ \|\ 
\mathrm{softmax}\!\left(\tfrac{z_{\text{stud}}}{\tau}\right)
\right),
\end{equation}
with $\tau=2.0$. For efficiency, we compute KL on
$U=\mathrm{TopK}_{\text{teach}}\cup \mathrm{TopK}_{\text{stud}}$
with $K=100$ and renormalize on $U$. We train with AdamW (lr $=10^{-5}$) for 5 epochs (effective batch size 32) and select checkpoints by dev TokJaccard@100. Intuitively, alignment encourages non-English queries to activate the same English planning evidence as their English counterparts while keeping the retrieval index unchanged.

%--------------------------------------
\begin{table*}[t]
\centering
% \caption{Effectiveness--efficiency trade-offs for PAG on MS~MARCO Dev (cf.\ Table~3 in the original paper) across planner
% set size $m$ (tokens/doc) and beam size $k$. We report MRR@10/Recall@10, index memory (GB), and latency (ms/query) broken
% into simultaneous (Simul.\ D.) and sequential (Seq.\ D.) decoding. Latency is hardware-dependent (80GB A100 in the
% original vs 96GB H100 here). \NA\ marks settings requiring unreleased artifacts; $^{\dagger}$ denotes reproduced
% effectiveness lower than reported.}
\caption{\textbf{RQ1: MS~MARCO Dev effectiveness--efficiency trade-offs} (cf.\ original Table~3) across planner set size
$m$ (tokens/doc) and beam size $k$. Latency values are hardware-specific (paper: A100 80GB; ours: H100 96GB).
\NA\ denotes settings requiring unreleased artifacts; $^{\dagger}$ marks reproduced effectiveness below the paper.}

\label{tab:rq1_efficiency}
\setlength{\tabcolsep}{4pt}
\begin{tabular}{cc cc cc cc cc cc}
\toprule
 &  & \multicolumn{2}{c}{MRR@10} 
 & \multicolumn{2}{c}{Recall@10} &
 \multicolumn{2}{c}{Index mem.\ (GB)} &
 \multicolumn{2}{c}{Simul.\ QL (ms)} &
 \multicolumn{2}{c}{Seq.\ QL (ms)} \\
\cmidrule(r){3-4} 
\cmidrule(r){5-6}
\cmidrule(r){7-8}
\cmidrule(r){9-10}
\cmidrule(r){11-12}
$m$ & $k$ & Reported & Reproduced & Reported & Reproduced & Reported & Computed & Reported & Computed & Reported & Computed \\
\midrule
16  & 10  & 0.342 & 0.330\rlap{$^{\dagger}$} & 0.577 & 0.556\rlap{$^{\dagger}$} & 1.30 & 1.13 & 20 & 4  & 44  & 42 \\
32  & 10  & 0.367 & 0.360\rlap{$^{\dagger}$} & 0.626 & 0.608\rlap{$^{\dagger}$} & 1.94 & 2.26 & 22 & 6  & 44  & 43 \\
64  & 10  & 0.379 & 0.380            & 0.641 & 0.644            & 3.27 & 4.53 & 25 & 8  & 44  & 46 \\
128 & 10  & 0.386 & \NA              & 0.645 & \NA              & 5.96 & \NA  & 31 & \NA & 44  & \NA \\
\midrule
16  & 100 & 0.355 & 0.341\rlap{$^{\dagger}$} & 0.620 & 0.606\rlap{$^{\dagger}$} & 1.30 & 1.13 & 20 & 4  & 250 & 262 \\
32  & 100 & 0.372 & 0.368\rlap{$^{\dagger}$} & 0.652 & 0.644\rlap{$^{\dagger}$} & 1.94 & 2.26 & 22 & 6  & 250 & 263 \\
64  & 100 & 0.385 & 0.386             & 0.670 & 0.671             & 3.27 & 4.53 & 25 & 8  & 250 & 266 \\
128 & 100 & 0.390 & \NA               & 0.664 & \NA               & 5.96 & \NA  & 31 & \NA & 250 & \NA \\
\bottomrule
\end{tabular}
\end{table*}

\section{Results and Analysis}
\label{sec:results}

We organize the empirical results around the three research questions introduced in Section~\ref{sec:intro}. We first assess whether PAG’s released artifacts suffice to reproduce its reported inference-time effectiveness and efficiency trends (RQ1). We then analyze how intent-preserving query variation affects planning stability and downstream ranking using the plan-drift diagnostics from Section~\ref{subsec:metrics} (RQ2). Finally, we evaluate fixed-index cross-lingual query shift and compare query-side mitigation strategies that preserve the released English index and document-side artifacts (RQ3).
\subsection{RQ1: Inference-time reproducibility}
\label{sec:rq1}

RQ1 evaluates inference-time reproducibility of PAG using released artifacts  under the
reported decoding configuration, focusing on effectiveness, decoding-time trade-offs, and inference-time ablations.

%---------------------------------------
\begin{table*}[t]
\centering
\caption{\textbf{Ablation on MS~MARCO Dev} (cf.\ Table~2 in the original paper). \NA\ denotes variants requiring
retraining or unreleased checkpoints.}
\label{tab:rq1_ablation}
\setlength{\tabcolsep}{4pt}
\begin{tabular}{l cc cc cc}
\toprule
 & \multicolumn{2}{c}{MRR@10} 
 & \multicolumn{2}{c}{Recall@10} 
 & \multicolumn{2}{c}{Index mem.\ (GB)} \\
\cmidrule(r){2-3}
\cmidrule(r){4-5}
\cmidrule(r){6-7}
Variant & Reported & Reproduced & Reported & Reproduced & Reported & Computed \\
\midrule
PAG & 0.385 & 0.386 & 0.670 & 0.671 & 3.27 & 4.53 \\
\midrule
1.\ w/o adding $s_{\text{simul}}(\cdot)$
    & 0.349 & 0.350 & 0.614 & 0.613 & 0.50 & 0.50 \\
2.\ Only $s_{\text{simul}}(\cdot)$ for retrieval
    & 0.303 & 0.303 & 0.569 & 0.569 & 2.77 & 2.77 \\
3.\ w/o seq2seq pre-training
    & 0.381 & \NA & 0.660 & \NA & 3.27 & \NA \\
4.\ w/o multi-obj.\ learning
    & 0.380 & \NA & 0.663 & \NA & 3.27 & \NA \\
\midrule
5.\ Only $M^{\text{set}}$
    & 0.322 & \NA & 0.606 & \NA & 2.77 & \NA \\
6.\ Only $M^{\text{seq}}$
    & 0.339 & \NA & 0.566 & \NA & 0.50 & \NA \\
7.\ Linear interp.\ of $M^{\text{set}}$ and $M^{\text{seq}}$
    & 0.360 & \NA & 0.593 & \NA & 3.27 & \NA \\
\midrule
8.\ Only $M^{sp}$
    & 0.378 & \NA & 0.667 & \NA & 35.28 & \NA \\
9.\ Only $M^{ds}$
    & 0.365 & \NA & 0.641 & \NA & 25.30 & \NA \\
\bottomrule
\end{tabular}
\vspace{2pt}
\end{table*}

\heading{Effectiveness}
Table~\ref{tab:rq1_effectiveness} compares reported effectiveness with results obtained using the released PAG checkpoint
and corpus-side artifacts at the default setting ($k=100$, $m=64$; cf.\ original Table~1). Across MS~MARCO Dev and
TREC-DL 2019/2020, reproduced PAG matches the reported values within 0.002 absolute difference (three-decimal precision),
indicating that the released artifacts suffice to reproduce the headline effectiveness in this setting. For context only
(i.e., neither part of the original PAG evaluation nor a reproduction target), we report results for three recent dense
retrievers spanning model scales~(Nomic-v2  ~\cite{nussbaum2025training}, EmbeddingGemma~\cite{vera2025embeddinggemma}, and Qwen3-8B~\cite{zhang2025qwen3}), next to the dense references from the
original paper, using a uniform brute-force scoring pipeline over the full corpus. These contextual baselines calibrate PAG's effectiveness and index-footprint trade-off relative to dense retrieval across model scales.

\heading{Efficiency and beam--latency trade-offs}
Table~\ref{tab:rq1_efficiency} characterizes PAG’s effectiveness--efficiency trends on MS~MARCO Dev (cf.\ original
Table~3) as a function of beam size $k$ and planner set size $m$. At the artifact-supported setting $m=64$, reproduced
effectiveness matches the reported values at both $k=10$ and $k=100$ (MRR@10 within 0.001; Recall@10 within 0.003).
Increasing beam size $k$ consistently improves effectiveness while substantially increasing sequential-decoding latency (Seq.\ QL $46.8\!\rightarrow\!268.6$\,ms for $k=10\!\rightarrow\!100$), matching the original qualitative trade-off. However, absolute efficiency values for simultaneous decoding and index memory differ from the paper: at $m=64$ we measure Simul.\ QL $=8.3$\,ms (vs.\ 25\,ms reported) and Index Mem.\ $=4.53$\,GB (vs.\ 3.27\,GB).
We therefore emphasize \emph{within-table trends} over absolute cross-hardware comparisons.
% As in the original paper, increasing $k$ yields consistent gains but substantially increases sequential-decoding latency
% (Seq.\ QL 46$\rightarrow$266\,ms for $k=10\rightarrow100$), while simultaneous decoding remains a small overhead (8\,ms at
% $m=64$). Latency is hardware-dependent, and memory can vary with accounting conventions; therefore, emphasis is placed
% on within-table trends.
The release provides only the top-$64$ planner tokens per document, enabling direct evaluation at $m=64$; $m=128$ is
marked \NA. For $m\in\{16,32\}$, we report a pragmatic approximation by truncating the top-$64$ token lists.
% These lists appear to be ordered by decreasing planning weight; thus truncation retains the highest-weight planning tokens
% and is a reasonable proxy for smaller-$m$ settings.
The released lists are ordered by decreasing planning weight, so truncation retains the highest-weight planning tokens and provides a reasonable proxy for smaller-$m$ settings.
Under truncation, effectiveness at $m\in\{16,32\}$ is lower than reported, but qualitative trends in $m$ and $k$ are
preserved.
\newline
\hfill \break
\noindent \textbf{Ablations}
Table~\ref{tab:rq1_ablation} reproduces inference-time ablations achievable with the released checkpoint and fixed
identifier/trie artifacts (others are \NA). Reproduced values closely match the reported ones and isolate the role of
planning-ahead guidance: \emph{relative to our reproduced PAG scores}, removing the look-ahead term (w/o adding
$s_{\text{simul}}$) reduces effectiveness by 0.036 MRR@10 and 0.058 Recall@10, while planning-only retrieval (Only
$s_{\text{simul}}$) drops further (0.083 MRR@10, 0.102 Recall@10). Overall, these ablations are consistent with PAG’s
intended use of the planner as look-ahead guidance for finite-beam sequential decoding rather than as a standalone
retriever. 
%Overall, the released artifacts reproduce PAG's headline effectiveness and qualitative efficiency trends, with remaining differences mainly due to hardware and unsupported larger-$m$ settings.

% ----------------------------
% Table 1: Retrieval quality (RQ2) -- compact format: score (Δ)
% ----------------------------
\begin{table*}[t]
\centering

% \caption{\textbf{RQ2: Retrieval performance under query variations.}
% Stage~1 (\emph{SimulOnly}) ranks by $s_{\mathrm{simul}}(q,d)$; Stage~2 runs full PAG with guided trie-constrained beam search.
% We report $\mu \pm \sigma$ across five perturbation seeds, with degradation in parentheses $\Delta = M(q)-M(\tilde q)$ (positive $\Delta$ indicates worse).
% Primary metrics are MRR@10 (MS~MARCO Dev) and NDCG@10 (TREC-DL 2019/2020).}

\caption{\textbf{RQ2: Retrieval under query variations.}
Stage~1 ranks by $s_{\mathrm{simul}}$; Stage~2 is full PAG.
Values are $\mu\pm\sigma$ across five seeds, with degradation $\Delta=M(q)-M(\tilde q)$ in parentheses.
Primary metrics: MRR@10 (Dev) and NDCG@10 (DL19/20).}

\label{tab:rq2-retrieval}
\setlength{\tabcolsep}{3pt}
\small
\begin{tabular}{ll c c c c}
\toprule
& & \multicolumn{2}{c}{\textbf{Stage 1: SimulOnly (Planner)}} &
    \multicolumn{2}{c}{\textbf{Stage 2: PAG (End-to-end)}} \\
\cmidrule(lr){3-4} \cmidrule(lr){5-6}
\textbf{Split} & \textbf{variation}
  & \textbf{NDCG@10 ($\Delta$)} & \textbf{MRR@10 ($\Delta$)}
  & \textbf{NDCG@10 ($\Delta$)} & \textbf{MRR@10 ($\Delta$)} \\
\midrule
\multirow{6}{*}{DL19}
  & Clean      
  & 0.643 $\pm$ 0.000 (--) & 0.898 $\pm$ 0.000 (--)
  & 0.669 $\pm$ 0.000 (--) & 0.915 $\pm$ 0.000 (--) \\

  & Misspelling  
  & 0.407 $\pm$ 0.020 (0.236 $\pm$ 0.020) & 0.637 $\pm$ 0.059 (0.260 $\pm$ 0.059)
  & 0.452 $\pm$ 0.012 (0.217 $\pm$ 0.012) & 0.726 $\pm$ 0.034 (0.189 $\pm$ 0.034) \\

  & Reordering     
  & 0.628 $\pm$ 0.013 (0.015 $\pm$ 0.013) & 0.899 $\pm$ 0.015 (-0.002 $\pm$ 0.015)
  & 0.654 $\pm$ 0.009 (0.014 $\pm$ 0.009) & 0.899 $\pm$ 0.007 (0.016 $\pm$ 0.007) \\

  & Synonymizing    
    & 0.454 $\pm$ 0.037 (0.189 $\pm$ 0.037) & 0.657 $\pm$ 0.040 (0.241 $\pm$ 0.040)
  & 0.526 $\pm$ 0.018 (0.143 $\pm$ 0.018) & 0.786 $\pm$ 0.023 (0.129 $\pm$ 0.023) \\

  & Paraphrasing 
    & 0.557 $\pm$ 0.030 (0.086 $\pm$ 0.030) & 0.798 $\pm$ 0.048 (0.099 $\pm$ 0.048)
  & 0.596 $\pm$ 0.012 (0.073 $\pm$ 0.012) & 0.871 $\pm$ 0.014 (0.044 $\pm$ 0.014) \\

  & Naturalizing  
   & 0.638 $\pm$ 0.000 (0.005 $\pm$ 0.000) & 0.932 $\pm$ 0.000 (-0.034 $\pm$ 0.000)
  & 0.626 $\pm$ 0.000 (0.043 $\pm$ 0.000) & 0.869 $\pm$ 0.000 (0.046 $\pm$ 0.000) \\
\midrule
\multirow{6}{*}{DL20}
  & Clean      
   & 0.638 $\pm$ 0.000 (--) & 0.930 $\pm$ 0.000 (--)
  & 0.621 $\pm$ 0.000 (--) & 0.869 $\pm$ 0.000 (--) \\

  & Misspelling 
  & 0.420 $\pm$ 0.015 (0.217 $\pm$ 0.015) & 0.678 $\pm$ 0.029 (0.252 $\pm$ 0.029)
  & 0.461 $\pm$ 0.019 (0.161 $\pm$ 0.019) & 0.703 $\pm$ 0.040 (0.166 $\pm$ 0.040) \\

  & Reordering     
  & 0.627 $\pm$ 0.004 (0.010 $\pm$ 0.004) & 0.919 $\pm$ 0.012 (0.010 $\pm$ 0.012)
  & 0.607 $\pm$ 0.009 (0.014 $\pm$ 0.009) & 0.848 $\pm$ 0.012 (0.021 $\pm$ 0.012) \\

  & Synonymizing    
    & 0.480 $\pm$ 0.042 (0.158 $\pm$ 0.042) & 0.711 $\pm$ 0.066 (0.219 $\pm$ 0.066)
  & 0.508 $\pm$ 0.007 (0.114 $\pm$ 0.007) & 0.741 $\pm$ 0.034 (0.129 $\pm$ 0.034) \\

  & Paraphrasing
    & 0.515 $\pm$ 0.017 (0.123 $\pm$ 0.017) & 0.777 $\pm$ 0.014 (0.152 $\pm$ 0.014)
  & 0.512 $\pm$ 0.019 (0.109 $\pm$ 0.019) & 0.738 $\pm$ 0.039 (0.131 $\pm$ 0.039) \\

  & Naturalizing
    & 0.615 $\pm$ 0.000 (0.023 $\pm$ 0.000) & 0.945 $\pm$ 0.000 (-0.016 $\pm$ 0.000)
  & 0.598 $\pm$ 0.000 (0.023 $\pm$ 0.000) & 0.869 $\pm$ 0.000 (0.000 $\pm$ 0.000) \\
\midrule
\multirow{6}{*}{Dev}
  & Clean       
  & 0.364 $\pm$ 0.000 (--) & 0.315 $\pm$ 0.000 (--)
  & 0.410 $\pm$ 0.000 (--) & 0.362 $\pm$ 0.000 (--) \\
  
  & Misspelling 
    & 0.220 $\pm$ 0.002 (0.145 $\pm$ 0.002) & 0.190 $\pm$ 0.002 (0.125 $\pm$ 0.002)
  & 0.245 $\pm$ 0.003 (0.165 $\pm$ 0.003) & 0.215 $\pm$ 0.003 (0.147 $\pm$ 0.003) \\
  
  & Reordering   
  
  & 0.355 $\pm$ 0.001 (0.009 $\pm$ 0.001) & 0.307 $\pm$ 0.001 (0.008 $\pm$ 0.001)
  & 0.399 $\pm$ 0.001 (0.011 $\pm$ 0.001) & 0.350 $\pm$ 0.001 (0.012 $\pm$ 0.001) \\
  
  & Synonymizing    

    & 0.260 $\pm$ 0.001 (0.104 $\pm$ 0.001) & 0.225 $\pm$ 0.001 (0.090 $\pm$ 0.001)
  & 0.305 $\pm$ 0.002 (0.105 $\pm$ 0.002) & 0.268 $\pm$ 0.002 (0.094 $\pm$ 0.002) \\
  
  & Paraphrasing
    & 0.297 $\pm$ 0.001 (0.068 $\pm$ 0.001) & 0.257 $\pm$ 0.001 (0.059 $\pm$ 0.001)
  & 0.342 $\pm$ 0.003 (0.069 $\pm$ 0.003) & 0.300 $\pm$ 0.002 (0.062 $\pm$ 0.002) \\
  & Naturalizing 
    & 0.341 $\pm$ 0.000 (0.023 $\pm$ 0.000) & 0.294 $\pm$ 0.000 (0.021 $\pm$ 0.000)
  & 0.388 $\pm$ 0.000 (0.022 $\pm$ 0.000) & 0.342 $\pm$ 0.000 (0.020 $\pm$ 0.000) \\
\bottomrule
\end{tabular}
\end{table*}

% ----------------------------
% Table 2: Plan stability + plan sensitivity (RQ2) -- compact format: MRR/NDCG in one cell
% ----------------------------
\begin{table*}[t]
\centering

% \caption{\textbf{RQ2: Planner stability and plan swapping.}
% \textbf{Stability:} CandOverlap@100 $=|D_{100}(q)\cap D_{100}(\tilde q)|/100$; TokJaccard@100 is Jaccard over top-100 planner tokens.
% \textbf{Sensitivity:} SeqGain is Stage~2 minus Stage~1 on $\tilde q$. PlanSwapDrop swaps in the clean plan ($q\!\rightarrow\!\tilde q$) and reports
% $M_{\text{PAG}}(\tilde q)-M_{\text{PAG}}(q\!\rightarrow\!\tilde q)$; negative implies the clean plan helps.
% Values are mean$\pm$std over five seeds.}

\caption{Planner stability and plan swapping (defined in \S\ref{subsec:metrics}). Mean$\pm$std over five seeds; PlanSwapDrop $<0$ implies the clean plan helps.}

\label{tab:rq2-collapse}
\setlength{\tabcolsep}{5pt}
\small
\begin{tabular}{ll c c c c}
\toprule
& & \multicolumn{2}{c}{\textbf{Plan Stability} ($\uparrow$)} &
\multicolumn{2}{c}{\textbf{Plan Sensitivity / Gain}} \\
\cmidrule(lr){3-4} \cmidrule(lr){5-6}
\textbf{Split} & \textbf{Perturbation}
  & \textbf{CandOverlap@100} & \textbf{TokJaccard@100}
  & \textbf{SeqGain (MRR/NDCG)} & \textbf{PlanSwapDrop (MRR/NDCG)} \\
\midrule
\multirow{5}{*}{DL19}
  & Misspelling
    & 0.374 $\pm$ 0.018 & 0.352 $\pm$ 0.015
  & 0.089 $\pm$ 0.035 / 0.045 $\pm$ 0.014
  & -0.070 $\pm$ 0.028 / -0.058 $\pm$ 0.011 \\

  & Ordering    
    & 0.801 $\pm$ 0.015 & 0.683 $\pm$ 0.019
  & -0.000 $\pm$ 0.014 / 0.026 $\pm$ 0.011
  & 0.000 $\pm$ 0.000 / -0.004 $\pm$ 0.004 \\

  & Synonym    
    & 0.439 $\pm$ 0.021 & 0.422 $\pm$ 0.023
  & 0.130 $\pm$ 0.019 / 0.071 $\pm$ 0.023
  & -0.033 $\pm$ 0.014 / -0.026 $\pm$ 0.008 \\
  
  & Paraphrase  
    & 0.625 $\pm$ 0.018 & 0.551 $\pm$ 0.009
  & 0.073 $\pm$ 0.040 / 0.039 $\pm$ 0.027
  & -0.015 $\pm$ 0.010 / -0.016 $\pm$ 0.010 \\
  
  & Naturality  
    & 0.738 $\pm$ 0.000 & 0.636 $\pm$ 0.000
  & -0.063 $\pm$ 0.000 / -0.012 $\pm$ 0.000
  & 0.000 $\pm$ 0.000 / -0.004 $\pm$ 0.000 \\
\midrule
\multirow{5}{*}{DL20}
  & Misspelling   
  & 0.348 $\pm$ 0.018 & 0.363 $\pm$ 0.014
  & 0.025 $\pm$ 0.054 / 0.041 $\pm$ 0.019
  & -0.013 $\pm$ 0.022 / -0.017 $\pm$ 0.016 \\

  & Ordering   
   & 0.801 $\pm$ 0.004 & 0.684 $\pm$ 0.008
  & -0.071 $\pm$ 0.018 / -0.020 $\pm$ 0.011
  & 0.006 $\pm$ 0.005 / 0.004 $\pm$ 0.004 \\
  & Synonym    
  & 0.498 $\pm$ 0.053 & 0.485 $\pm$ 0.032
  & 0.030 $\pm$ 0.080 / 0.028 $\pm$ 0.045
  & -0.024 $\pm$ 0.017 / -0.018 $\pm$ 0.007 \\
  & Paraphrase  
    & 0.569 $\pm$ 0.021 & 0.522 $\pm$ 0.014
  & -0.039 $\pm$ 0.032 / -0.003 $\pm$ 0.010
  & -0.025 $\pm$ 0.008 / -0.011 $\pm$ 0.003 \\
  & Naturality   
  & 0.708 $\pm$ 0.000 & 0.577 $\pm$ 0.000
  & -0.076 $\pm$ 0.000 / -0.017 $\pm$ 0.000
  & -0.000 $\pm$ 0.000 / -0.008 $\pm$ 0.000 \\
\midrule
\multirow{5}{*}{Dev}
  & Misspelling
    & 0.312 $\pm$ 0.002 & 0.336 $\pm$ 0.002
  & 0.025 $\pm$ 0.003 / 0.026 $\pm$ 0.003
  & -0.015 $\pm$ 0.001 / -0.016 $\pm$ 0.001 \\

  & Ordering   
    & 0.767 $\pm$ 0.002 & 0.661 $\pm$ 0.002
  & 0.043 $\pm$ 0.001 / 0.044 $\pm$ 0.001
  & -0.001 $\pm$ 0.000 / -0.001 $\pm$ 0.000 \\
  & Synonym    
    & 0.460 $\pm$ 0.002 & 0.475 $\pm$ 0.001
  & 0.042 $\pm$ 0.002 / 0.045 $\pm$ 0.002
  & -0.014 $\pm$ 0.000 / -0.015 $\pm$ 0.000 \\
  & Paraphrase  
    & 0.573 $\pm$ 0.002 & 0.536 $\pm$ 0.002
  & 0.043 $\pm$ 0.002 / 0.045 $\pm$ 0.003
  & -0.007 $\pm$ 0.000 / -0.007 $\pm$ 0.001 \\
  & Naturality  
    & 0.680 $\pm$ 0.000 & 0.584 $\pm$ 0.000
  & 0.048 $\pm$ 0.000 / 0.048 $\pm$ 0.000
  & -0.001 $\pm$ 0.000 / -0.001 $\pm$ 0.000 \\
\bottomrule
\end{tabular}
\end{table*}

\subsection{RQ2: Robustness stress test}
\label{sec:rq2}

\begin{figure}[t]
  \centering
  \includegraphics[width=\columnwidth]{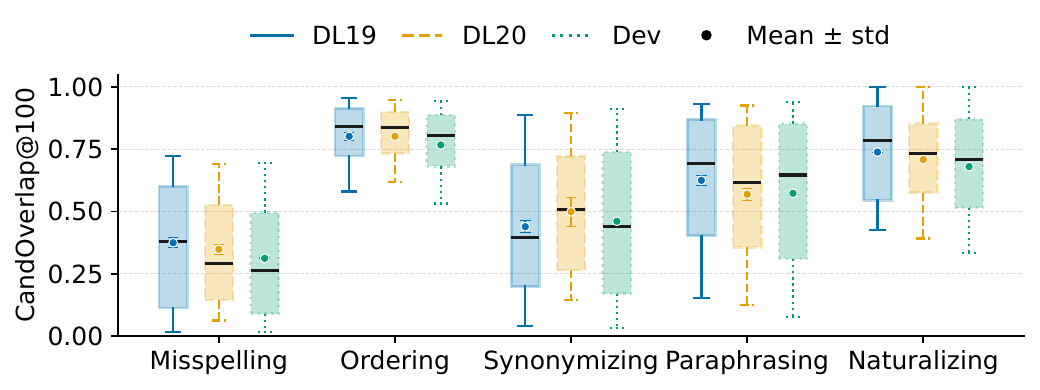}
%   \caption{ Candidate-set stability (CandOverlap@100).
% CandOverlap@100 is overlap@100 between the planner top-100 candidate sets for clean $q$ and perturbed $\tilde q$.
% Across five seeds: whiskers $p10$--$p90$, boxes $p25$--$p75$, center line median, marker mean $\pm$ std (Naturality: std $=0$).}
\caption{Candidate-set stability (CandOverlap@100).
CandOverlap@100 compares the planner top-100 candidate sets for clean $q$ vs.\ perturbed $\tilde q$.
Line styles denote splits (DL19 solid, DL20 dashed, Dev dotted).
Across five seeds: whiskers $p10$--$p90$, boxes $p25$--$p75$, median line, mean $\pm$ std marker (Naturality: std $=0$).}

  \label{fig:rq2-candoverlap}
  \vspace{-1mm}
\end{figure}

\begin{figure}[t]
  \centering
  \includegraphics[width=\columnwidth]{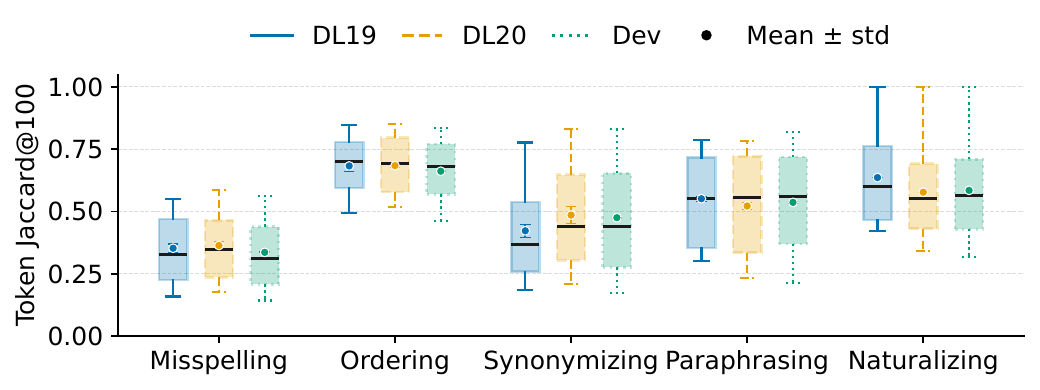}
%   \caption{\textbf{RQ2: Planner-token stability (TokJaccard@100).}
% TokJaccard@100 is Jaccard similarity between the planner top-100 tokens for clean $q$ and perturbed $\tilde q$.
% Line styles denote splits (DL19 solid, DL20 dashed, Dev dotted).
% Across five seeds: whiskers $p10$--$p90$, boxes $p25$--$p75$, center line median, marker mean $\pm$ std (Naturality: std $=0$).}
\caption{Planner-token stability (TokJaccard@100).
TokJaccard@100 compares top-100 planner tokens for clean $q$ vs.\ perturbed $\tilde q$; same plotting convention as Fig.~\ref{fig:rq2-candoverlap}.}

  \label{fig:rq2-tokjaccard}
  \vspace{-1mm}
\end{figure}

RQ2 stress-tests PAG under intent-preserving query variation while holding the released checkpoint, docids, trie, and
decoding configuration fixed. This isolates query-side sensitivity in Stage~1 planning (\emph{SimulOnly}, ranking by
$s_{\mathrm{simul}}$) and in Stage~2 planning-guided decoding.
\newline
\hfill
\noindent \textbf{Query-variation evaluation.}
We evaluate five intent-preserving variations of each clean query $q$.  Table~\ref{tab:rq2-retrieval} shows that lexical variations cause large end-to-end degradation, while reordering is near-invariant.
On DL19, Stage~2 drops by 0.217 NDCG (misspelling), 0.143 (synonym), and 0.073 (paraphrase), versus 0.014 for reordering; DL20 shows the same ranking
(0.161/0.114/0.109 vs.\ 0.014). These effectiveness gaps align with plan stability in Table~\ref{tab:rq2-collapse}:
reordering has the highest overlap (CandOverlap@100 $\approx$0.80; TokJaccard@100 $\approx$0.68 on DL19/20), whereas misspelling and synonym sharply
reduce overlap (e.g., DL19 CandOverlap 0.374/0.439; TokJaccard 0.352/0.422; DL20 CandOverlap 0.348/0.498; TokJaccard 0.363/0.485), and the boxplots
(Fig.~\ref{fig:rq2-candoverlap}--\ref{fig:rq2-tokjaccard}) show pronounced lower tails under lexical corruption, indicating that a non-trivial subset of
queries undergoes near-replacement of the planned set. A plausible contributor to plan collapse under typos is \emph{subword fragmentation}:
small edits can change SentencePiece segmentation, producing rare/mismatched units that fail to fire the lexical planner’s sparse triggers and sharply reduce
planned-set overlap.
%
% CandOverlap figure
%
% TokJaccard figure
Consistent with PAG’s mechanism, this instability limits Stage~2 because the look-ahead bonus is computed
from a shifted candidate pool, reducing coverage of relevant docid prefixes and increasing pruning risk. Stage~2 improvements are therefore conditional:
SeqGain is positive for most Dev conditions (0.043--0.048) and for DL19 under misspelling/synonym/paraphrase (0.045/0.071/0.039 NDCG), but becomes negative
in several DL20 settings despite relatively high overlap (e.g., Reordering CandOverlap 0.801 and TokJaccard 0.684, yet SeqGain $=-0.020$ NDCG), indicating
residual sensitivity in the sequential component beyond plan membership. 
\\
Finally, plan swapping isolates causality: PlanSwapDrop is typically negative under
harder variations (e.g., DL19 misspelling $-0.058$ NDCG and $-0.070$ MRR), showing that reusing the clean bonus partially recovers performance, but the
magnitude is smaller than the total losses in Table~\ref{tab:rq2-retrieval}, implying that robustness is bounded by both planning instability (especially
tail events) and sequential-decoding sensitivity under perturbed inputs.

\vspace*{-2mm}
\heading{Plan-collapse}
Table~\ref{tab:plan_collapse} reports seed-aggregated collapse rates over 75 (split, variation, seed) conditions
(176{,}925 query instances). Collapse is infrequent on MS~MARCO Dev (3.2--6.2\%) but higher on TREC-DL
(2.3--11.6\% on DL19; 9.6--11.1\% on DL20), indicating a non-trivial subset of queries with both low stability and a
planning-only drop ($\Delta M_{\text{SimulOnly}}\le-0.05$).
Ordering exhibits a large low-stability tail (19.0--33.3\%; Table~\ref{tab:rq2-collapse}) yet comparatively lower collapse, suggesting that overlap
deviations under ordering often do not coincide with large planning-only drops at $\delta=0.05$.
By contrast, misspelling/synonym/paraphrase yield 9.6--11.6\% collapse on TREC-DL, indicating tail events where planning becomes both unstable and
planning-only effectiveness degrades.
% \looseness=-1

% \looseness=-1

%______________________________________

\begin{figure*}[t]
  \centering
  \includegraphics[width=\textwidth]{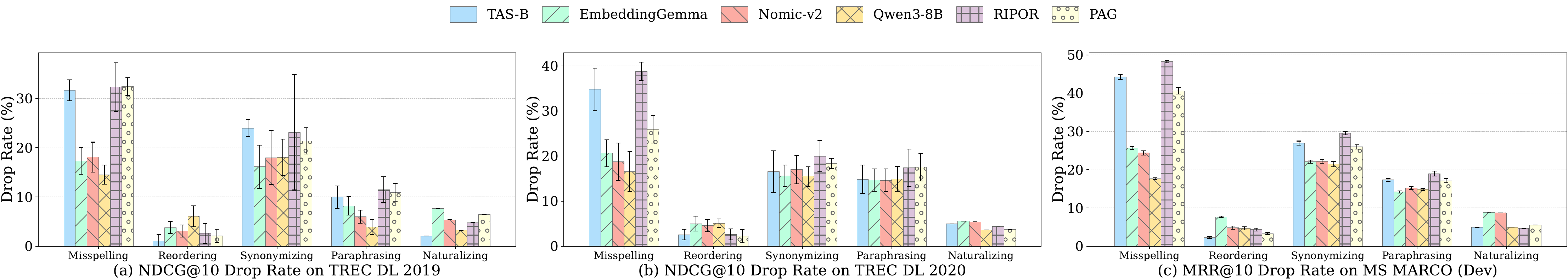}
\caption{\textbf{Robustness under query variations: PAG vs.\ dense and GR baselines.}
Mean relative drop rate (\%) under five intent-preserving query variations, computed relative to the original query.
We compare PAG and a strong GR baseline (RIPOR) against recent strong dense retrievers on MS~MARCO Dev (MRR@10)
and TREC-DL 2019/2020 (NDCG@10). Error bars indicate variability across queries.}
  \label{fig:dense_VS_pag}
  \vspace{-1mm}
\end{figure*}

\begin{table}[t]
% \caption{\textbf{Plan collapse analysis.} We report the \textbf{collapse rate} (percentage of query pairs satisfying the collapse condition) and the stability threshold $\tau$ (defined per condition as the 10th percentile of $\mathrm{CandOverlap}@100$). A query pair is collapsed if it exhibits low stability $\big(\mathrm{CandOverlap}@100 < \tau \ \lor\ \mathrm{TokJaccard}@100 < \tau\big)$ and a planning-only performance drop $\Delta M_{\text{SimulOnly}} \le -0.05$. Results are averaged across five seeds ($\mu \pm \sigma$). $\tau$ is computed per seed-condition (split $\times$ variation $\times$ seed); the table reports the seed-mean $\mathbb{E}[\tau]$ (std.\ omitted for brevity).}
\caption{\textbf{Plan collapse analysis.}
Collapse rate (mean$\pm$std over five seeds) under the criterion from \S\ref{subsec:metrics} ($\delta=0.05$, $\tau=\mathrm{p}10$ of CandOverlap@100 per seed-condition).
We also report the seed-mean threshold $\mathbb{E}[\tau]$ (std.\ omitted for brevity).}

    \label{tab:plan_collapse}
    \centering
    \resizebox{\columnwidth}{!}{
    \begin{tabular}{llcc}
        \toprule
        \textbf{Split} & \textbf{Variation} & \textbf{Collapse Rate (\%)} & \textbf{Threshold ($\tau$)} \\
        \midrule
        \multirow{5}{*}{MS MARCO Dev} 
        & Misspelling & $5.2 \pm 0.1$ & 0.015 \\
        & Ordering & $3.2 \pm 0.8$ & 0.531 \\
        & Synonym & $5.7 \pm 0.1$ & 0.032 \\
        & Paraphrase & $6.2 \pm 0.1$ & 0.078 \\
        & Naturality & $4.5 \pm 0.0$ & 0.333 \\
        \midrule
        \multirow{5}{*}{TREC-DL 2019} 
        & Misspelling & $11.2 \pm 1.0$ & 0.015 \\
        & Ordering & $7.4 \pm 3.0$ & 0.580 \\
        & Synonym & $11.6 \pm 0.0$ & 0.040 \\
        & Paraphrase & $10.7 \pm 1.3$ & 0.152 \\
        & Naturality & $2.3 \pm 0.0$ & 0.427 \\
        \midrule
        \multirow{5}{*}{TREC-DL 2020} 
        & Misspelling & $9.6 \pm 1.5$ & 0.063 \\
        & Ordering & $10.0 \pm 1.7$ & 0.617 \\
        & Synonym & $10.4 \pm 2.5$ & 0.144 \\
        & Paraphrase & $11.1 \pm 0.0$ & 0.126 \\
        & Naturality & $11.1 \pm 0.0$ & 0.392 \\
        \bottomrule
    \end{tabular}
    }  \vspace{-1mm}
\end{table}
%______________________________________
% \vspace*{-2mm}
\heading{Comparison with dense and GR baselines}
Fig.~\ref{fig:dense_VS_pag} shows that robustness is primarily limited by \emph{lexical disruption} rather than benign rewrites:
misspellings and synonym substitutions strongly affect both PAG/RIPOR and the dense baseline TAS-B, while recent dense retrievers are typically less impacted across
MS~MARCO Dev and TREC-DL 2019/2020.
Under misspellings, RIPOR/TAS-B/PAG drop by $\sim$48\%/44\%/41\% on MS~MARCO Dev, versus $\sim$18--26\% for recent dense models, and the same ordering holds on DL19/20.
Reordering is near-invariant for all methods (single-digit drops), indicating brittleness is driven by surface-form mismatch rather than word order.
Synonym substitution shows a similar pattern: recent dense models drop $\sim$16--22\%, while RIPOR/PAG typically drop $\sim$18--30\%, placing PAG closer to RIPOR than to recent dense retrievers under lexical perturbations.
Overall, this comparison aligns with RQ2--RQ3: robustness failures concentrate on lexical surface-form shifts, which in RQ2 coincide with planner drift and reduced candidate stability that can limit Stage~2 gains under finite-beam decoding.
RQ3 confirms the same sensitivity under stronger surface-form mismatch (cross-lingual queries) with a fixed index and evaluates query-side mitigations.

\subsection{RQ3: Cross-lingual query shift}
\label{sec:rq3}

RQ2 showed that lexical surface-form variation can destabilize the planning signal and weaken downstream decoding. RQ3 tests this failure mode under a more extreme setting: \emph{cross-lingual query shift}, where query surface form diverges sharply from the English planning vocabulary on which PAG was trained, while the document collection and all document-side artifacts remain fixed to the released English index (set-based identifiers $t_d$, sequential docids $c_d$, and the trie). This setting is \emph{not} studied in the original paper, so we report it as a stress test rather than a reproduction claim. It is also a direct test of PAG's original motivation: if the planner cannot recover a useful candidate pool under fixed-index mismatch, then the look-ahead bonus cannot protect relevant prefixes from early pruning.

\heading{Setting}
We issue mMARCO queries in four languages (nl, fr, de, zh) against the fixed English MS~MARCO passage collection (Section~\ref{subsec:crosslingual}), keeping decoding and all document-side artifacts unchanged. For zh, the released English tokenizer/vocabulary can tokenize characters lossily, so \emph{Naive} and \emph{Aligned} reflect both cross-lingual shift and vocabulary-coverage failure, while \emph{Translate} mitigates this by converting queries to English before tokenization.

We compare four query-side settings:
(i) \emph{Naive}, which applies the released English PAG pipeline directly to non-English queries;
(ii) \emph{Seq-only}, which disables planning guidance and decodes using only $s(c_{\le i};q)$;
(iii) \emph{Translate}, which translates queries into English at inference time and then runs the unchanged English PAG pipeline; and
(iv) \emph{Aligned}, which fine-tunes only query-side parameters to match the released English planner-token distribution on paired $(q^{\mathrm{en}}, q^\ell)$ queries, without re-indexing or modifying $t_d$, $c_d$, or the trie.

\heading{Headline result}
Under a fixed English index, naive cross-lingual transfer substantially degrades planning-guided decoding because non-English queries often fail to activate the English planner evidence that supplies the look-ahead bonus. Query translation provides the strongest recovery across languages, while planner-token alignment yields only partial gains because improvements in token overlap do not consistently translate into candidate-set alignment.

\heading{Result analysis}

\begin{figure}[t]
 \centering
  \includegraphics[width=\linewidth]{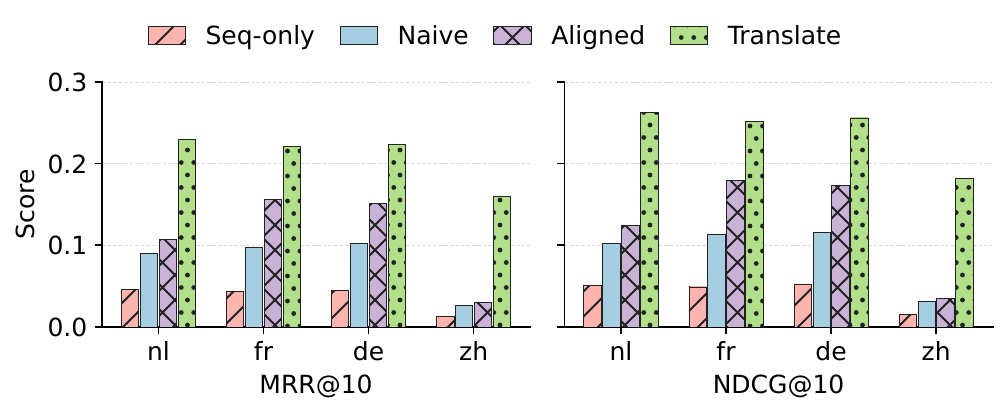}
\caption{\textbf{RQ3: Cross-lingual query shift.}
Stage~2 effectiveness (MRR@10, NDCG@10) on nl/fr/de/zh mMARCO queries with a fixed English index.
We compare \emph{Naive}, \emph{Translate}, and \emph{Aligned}; \emph{Seq-only} disables planning guidance.}

\label{fig:pipeline_stress_vertical}
\end{figure}

\noindent\textbf{Naive transfer.}
\emph{Naive} transfer yields low Stage~2 MRR@10 across languages (nl 0.090, fr 0.097, de 0.102, zh 0.027; Fig.~\ref{fig:pipeline_stress_vertical}) and coincides with weak agreement to the English reference (TokJaccard@100 0.072--0.102; CandOverlap@100 0.124--0.176; Table~\ref{tab:rq3-diagnostics}). This indicates that non-English queries often fail to activate the English planner-token evidence and top-100 planned candidates used to supply the look-ahead bonus.
% \looseness=-1

% \looseness=-1
\noindent\textbf{Translation.}
\emph{Translate} provides the strongest recovery (Stage~2 MRR@10: nl 0.230, fr 0.221, de 0.224, zh 0.160), improving over \emph{Naive} by +0.140/+0.124/+0.122/+0.133 MRR@10, respectively (Fig.~\ref{fig:pipeline_stress_vertical}). It also substantially increases overlap with the English reference (e.g., TokJaccard@100 nl 0.101$\rightarrow$0.430 and zh 0.072$\rightarrow$0.317; CandOverlap@100 nl 0.170$\rightarrow$0.563 and zh 0.124$\rightarrow$0.449; Table~\ref{tab:rq3-diagnostics}), consistent with translation restoring compatibility with the fixed English planning vocabulary and candidate coverage.

\noindent\textbf{Sequential-only ablation.}
Removing the planner prior is consistently harmful: \emph{Seq-only} is uniformly poor (Stage~2 MRR@10 0.013--0.046) and falls below end-to-end \emph{Naive} for every language (Fig.~\ref{fig:pipeline_stress_vertical}). This shows that trie-constrained decoding without planning guidance does not reliably traverse the fixed English docid space under cross-lingual inputs.

\noindent\textbf{Planner-token alignment.}
\emph{Aligned} yields partial gains for nl/fr/de in Stage~2 MRR@10 (0.090$\rightarrow$0.107, 0.097$\rightarrow$0.156, 0.102$\rightarrow$0.151), but is minimal for zh (0.027$\rightarrow$0.030) and remains far below \emph{Translate}. The diagnostics explain this ceiling: TokJaccard@100 and its lower tail improve for nl/fr/de, while CandOverlap@100 does not consistently increase (nl 0.170$\rightarrow$0.121, fr 0.176$\rightarrow$0.176, de 0.154$\rightarrow$0.171, zh 0.124$\rightarrow$0.029; Table~\ref{tab:rq3-diagnostics}). Thus, token-level planner alignment does not reliably translate into planned-set alignment.

\noindent\textbf{Residual mismatch.}
Residual gaps persist even under \emph{Translate}: zh remains below nl/fr/de in Stage~2 MRR@10 (0.160 vs.\ 0.221--0.230) and retains lower overlap (TokJaccard@100 0.317 vs.\ 0.402--0.430; CandOverlap@100 0.449 vs.\ 0.537--0.563; Table~\ref{tab:rq3-diagnostics}), consistent with remaining mismatch under a fixed English index.

Overall, under a fixed English index, Stage~2 degrades when planning fails to recover English planner evidence, causing the look-ahead bonus to rely on an irrelevant candidate pool. Translation restores overlap and recovers most performance without re-indexing, whereas token-level alignment yields only partial gains because candidate-set overlap does not consistently improve.
% ----------------------------
% RQ3: Main effectiveness table (fill values)
% ----------------------------
% \input{tables/RQ3-cross}

% ----------------------------
% RQ3: Planner diagnostics table (fill values)
% ----------------------------
\begin{table}[t]
\centering
\caption{\emph{RQ3: Planner overlap diagnostics under cross-lingual query shift.}
Overlap between each $q^{\ell}$ and its English reference $q^{\mathrm{en}}$ (same query id). p10(TokJaccard@100) reports the 10th-percentile (lower-tail) token overlap.}
\label{tab:rq3-diagnostics}
\setlength{\tabcolsep}{4pt}
\small
\begin{tabular}{l l ccc}
\toprule
\textbf{Language} & \textbf{Diagnostic}
& \textbf{Naive} & \textbf{Aligned} & \textbf{Translate} \\
\midrule
\multirow{3}{*}{\emph{Dutch (nl)}}
& TokJaccard@100        & 0.101 & 0.134 & 0.430 \\
& CandOverlap@100       & 0.170 & 0.121 & 0.563 \\
& p10(TokJaccard@100)   & 0.020 & 0.042 & 0.124 \\

\midrule
\multirow{3}{*}{\emph{French (fr)}}
& TokJaccard@100        & 0.102 & 0.182 & 0.402 \\
& CandOverlap@100       & 0.176 & 0.176 & 0.537 \\
& p10(TokJaccard@100)   & 0.031 & 0.070 & 0.117 \\

\midrule
\multirow{3}{*}{\emph{German (de)}}
& TokJaccard@100        & 0.090 & 0.194 & 0.408 \\
& CandOverlap@100       & 0.154 & 0.171 & 0.543 \\
& p10(TokJaccard@100)   & 0.026 & 0.070 & 0.124 \\

\midrule
\multirow{3}{*}{\emph{Chinese (zh)}}
& TokJaccard@100        & 0.072 & 0.082 & 0.317 \\
& CandOverlap@100       & 0.124 & 0.029 & 0.449 \\
& p10(TokJaccard@100)   & 0.020 & 0.020 & 0.087 \\
\bottomrule
\end{tabular}
\end{table}

\section{Conclusion}
We conducted an inference-time reproduction and stress-test study of Planning Ahead in Generative Retrieval (PAG).
For \emph{RQ1}, using the released checkpoint and corpus-side artifacts under the reported decoding configuration, we reproduced the headline effectiveness results on MS~MARCO Dev and TREC-DL 2019/2020 and corroborated the qualitative beam--latency trade-off.
For \emph{RQ2}, we instrumented the planning stage under intent-preserving query variation and showed that lexical perturbations, such as misspellings and synonym substitutions, can substantially shift the planner's top-$n$ candidate set and high-weight planner tokens.
These plan-drift effects coincide with reduced candidate coverage and weaker end-to-end ranking, consistent with increased pruning risk under finite-beam decoding.
More broadly, the results indicate that PAG's planning signal is tightly coupled to query surface form: when lexical variation shifts the planned candidate pool, the look-ahead bonus becomes less informative and can in some cases collapse.
A plausible contributor is subword fragmentation, whereby small edits alter SentencePiece segmentation and suppress the sparse lexical evidence used to activate planned documents.
For \emph{RQ3}, we evaluated cross-lingual query shift under a fixed English identifier space and found that performance degrades markedly under language mismatch.
Among the query-side mitigations we tested, query translation provides the strongest recovery, while lightweight planner-token alignment improves over naive cross-lingual use but remains limited without translation.

Taken together, our results show that planning-guided decoding is reproducible and effective under the released inference setup, but its gains depend on the stability of the planning signal under realistic variation and shift.

\heading{Takeaways for planning-guided GR}
Our findings suggest three practical lessons for future work on planning-guided decoding in GR.
First, \emph{planner robustness is a first-order design concern}: surface-form sensitivity is not merely an auxiliary evaluation issue, because under lexical perturbation the planning bonus can weaken enough to approach unguided beam search.
Second, \emph{plan drift is diagnostically informative}: candidate-set and planner-token overlap expose failure modes that are not visible from end-to-end effectiveness alone and should be reported alongside ranking metrics in future robustness evaluations.
Third, \emph{translation is a strong no-reindex baseline under language mismatch}: in our fixed-index setting, simple query translation consistently outperforms lightweight planner-token alignment, suggesting that restoring compatibility with the planner's evidence space is more effective than token-level alignment alone.

\heading{Limitations and future work}
This study is bounded by the released inference artifacts (checkpoint, identifiers, trie, and top-$m$ planner tokens), and we cannot evaluate training-stage variants or settings requiring unreleased intermediate artifacts.
Latency measurements depend on our hardware and should therefore be interpreted as relative trends rather than directly comparable absolute values.
Our stress tests also keep the English index fixed, isolating query-side shift, including language mismatch, rather than corpus-side drift or multilingual document collections.
Future work should evaluate (i) planner robustness under corpus-side drift and alternative identifier or trie constructions, (ii) stronger query-side adaptation strategies beyond translation and token distillation, such as multilingual planning signals or jointly trained planners, and (iii) robustness protocols that jointly report end-to-end metrics and intermediate diagnostics, such as candidate coverage, plan drift, and tail-risk indicators, across datasets and beam regimes.

\heading{Reproducibility}
Our code and artifacts are available at \url{https://github.com/kidist-amde/lost-in-decoding}.
\begin{acks}
This research was supported by the Dutch Research Council (NWO), under project numbers 024.004.022, NWA.1389.20.\-183, and KICH3.\-LTP.20.006, the European Union under grant agreement No. 101201\-510 (UNITE), the China Scholarship Council (202308440220), Swiss National Science Foundation, grant 215742. Views and opinions expressed are those of the author(s) only and do not necessarily reflect those of their respective employers, funders, and/or granting authorities.    
\end{acks}

\balance
\bibliographystyle{ACM-Reference-Format}
\bibliography{main}

\end{document}